\documentclass[journal=jacsat,manuscript=article]{achemso}

\usepackage[version=3]{mhchem} 
\usepackage{amsmath}
\usepackage{xcolor}
\newcommand{\jb}[1]{\textcolor{black}{#1}} 

\author{Julien Bauland}
\affiliation{Department of Materials, ETH Zurich, Vladimir-Prelog-Weg 5, Zurich, 8093, Switzerland}
\author{Tim J. Wooster}
\affiliation{Nestlé Institute of Food Sciences, Nestlé Research, Route de Jorat 57, Vers Chez les Blanc, 1000 Lausanne 26, Switzerland}
\author{Peter Fischer}
\affiliation{Institute of Food, Nutrition, and Health, ETH Zurich, Schmelzbergstrasse 7, Zürich, 8092, Switzerland}
\author{Jan Vermant}
\affiliation{Department of Materials, ETH Zurich, Vladimir-Prelog-Weg 5, Zurich, 8093, Switzerland}
\email{jan.vermant@mat.ethz.ch}

\title[An \textsf{achemso} demo]
 {Thermo-Rheological Memory of $\kappa$-Carrageenan Fluid Gels Formed Under Flow}


\begin{document}


\begin{abstract}

Fluid gels are soft materials formed by shearing biopolymer solutions during  the sol–gel transition. Their ability to  yield and flow beyond a critical stress makes them attractive for designing versatile, biocompatible materials in food, health care and medical applications. Although it is well established that both microstructure and mechanical properties depend on the shear applied during gelation, a unified physical framework linking these features remains lacking. Here, using $\kappa$-carrageenan gels as a model system, we use a combination of rheology and confocal microscopy to tackle their shear-induced structuring in fluid gels. We identify a thermo-rheological memory in $\kappa$-carrageenan gels formed under flow and show that it arises from a competition between shear and interparticle adhesion, captured by an Adhesion number. The resulting microstructural evolution is reminiscent of the behavior of attractive particulate dispersions under simple shear flow, thereby bridging gels made of macromolecules and particulate gels. This framework provides a route to tune fluid gel properties without altering their composition.

\end{abstract}




\section{Introduction}

Fluid gels are weak solids obtained by shearing a polymer solution as it is undergoing a sol–gel transition. Traditionally, the term ``fluid gels'' refers to biopolymer hydrogels based on hydrocolloids or proteins, such as carrageenan~\cite{Gabriele2009,Hilliou2007}, gellan gum~\cite{Sworn1995,Caggioni2007}, methylcellulose~\cite{Nelson2022}, alginate~\cite{Farrs2014}, or gelatin~\cite{DeCarvalho1997}, with applications in food and cosmetic formulations~\cite{Wooster2023}. In such systems, the sol–gel transition is typically governed by temperature, which controls the self-organization of the polymer chains. Under the influence of shearing during the gelation, fluid gels acquire an heterogeneous, dispersion microstructure consisting of hydrated polymeric particles with sizes ranging from 1 to 100~\textmu m~\cite{DOria2023a}, \jb{placing them in the broad category of particulate gels}. 
This heterogeneous granular-like assembly contrasts with the homogeneous polymer network obtained from the same composition under quiescent conditions~\cite{Caggioni2007}. 

The functionality of fluid gels arises from their ability to sustain elastic stresses at rest while yielding and flow once a critical stress is exceeded (for a given time-frame), placing them among yield stress fluids~\cite{Nelson2019}. Below the yield stress, fluid gels behave as soft solids with a finite shear modulus, while above it, microstructural rearrangement enable viscous flow, typically accompanied by shear thinning. This solid–liquid duality underpins their technological versatility: it provides texture and stability at rest while allowing the material to be handled, poured, or spread when stressed. Fluid gels are also designed to ensure the stability of multiphase formulations, such as dispersions of dense particles or emulsions. The presence of an elastic structure will  counteract gravitational and buoyant forces, and hence inhibit or delay sedimentation or creaming of dispersed phases.

Beyond the magnitude of the stress at the yielding transition, the microstructure of such fluid gels has been found to play a crucial role in preventing particle sedimentation and ensuring material performance~\cite{Emady2013}. In fluid gels, the arrested microstructure is widely understood to result from a balance between shear forces applied during gelation and attractive interactions between polymer chains~\cite{DeCarvalho1997,Frith2002}. Numerous studies have shown that key microstructural features, such as particle size and connectivity, as well as the resulting mechanical properties, can be tuned  by the imposed shear rate during gelation~\cite{Gabriele2009,Hilliou2007,DOria2024}, offering a processing based route to modulate material properties at constant composition. Yet, a clear understanding of how  the shearing speed determine fluid gel microstructure and associated mechanical behavior remains lacking~\cite{Wooster2023}.

Here, combining rheology and confocal microscopy, we investigate $\kappa$-carrageenan fluid gels formed under controlled shear rates during gelation. Using large-amplitude oscillatory shear (LAOS), we show that the shear rate during gelation is encoded as a rheological memory in the nonlinear viscoelastic response, while having only a weak influence on the linear viscoelasticity and yield stress. Strain-amplitude sweeps reveal that the $G^{\prime\prime}$ overshoot, associated with plastic flow, increases with the shear rate during gelation up to a critical value, above which it levels off. Confocal microscopy observations demonstrate that this macroscopic trend arises from shear-induced variations in microstructure: the average particle size increases with shear rate up to the same critical value and then remains constant. By varying KCl concentration, we show that this critical shear rate scales with the ionic strength of the solution. This behavior can be further rationalized using an Adhesion number ($Ad$), analogous to the Mason number ($Mn$), defined here as the ratio of the ion-mediated attractive stresses to the viscous stress imposed by the  background fluid. 

\section{Materials and Methods}
\label{sec:exp}

\subsection{Sample Preparation}

Samples are prepared by dispersing $0.35~\%~(w/w)$ $\kappa$-carrageenan powder (Sigma-Aldrich) at $80^{\circ}\mathrm{C}$ under magnetic stirring for 30~min. Potassium chloride (KCl, Sigma-Aldrich, $M_\mathrm{w} = 74.55~\mathrm{g/mol}$) is added at concentrations between 10 and 100~mM. To ensure stability, sodium azide (NaN\textsubscript{3}, Sigma-Aldrich, $M_\mathrm{w} = 65.01~\mathrm{g/mol}$) is added as a bio-preservative at a final concentration of $0.2~\mathrm{g/L}$, corresponding to an ionic strength increase of 3.1~mM. The solution is then cooled to room temperature under stirring and used within two weeks.

\subsection{Rheometry}

Rheological measurements were performed using a stress-controlled rheometer (MCR~702, Anton Paar) equipped with a coaxial cylinder geometry (CC27, inner radius $r_i = 26.66~\mathrm{mm}$, outer radius $r_e = 28.924~\mathrm{mm}$). A custom-built solvent trap was used; its design was inspired by Sato et al.~\cite{sato2005} and allows repeated heating cycles without measurable variations in the high-temperature viscosity of the carrageenan solutions. \jb{Carrageenan fluid gel production and characterization involved loading the solutions into the sealed geometry and heating to $70^{\circ}$C. A cooling ramp was then performed from 70 to $25^{\circ}$C at $0.05^{\circ}$C/s under a constant shear rate $\dot{\gamma}_0$. This shear rate was maintained at $25^{\circ}$C for 120~s to equilibrate the system. Immediately following, a flow curve was measured by ramping the shear rate from 1000 down to $1~\rm{s}^{-1}$, and ramping up again, with an averaging time of 10~s per point. Upon flow cessation ($\dot{\gamma}=0$), small-amplitude oscillatory shear (SAOS) experiments were typically performed at a strain amplitude $\gamma = 3.10^{-2}$ and an angular frequency of $0.16~\rm{rad.s^{-1}}$.}

\subsection{Confocal Microscopy}
\label{sec:conf}
Dispersion of polysaccharides such as carrageenans provides limited optical contrast between polymer assemblies and solvent, preventing the use of bright-field imaging. For confocal microscopy, the lack of a simple non-specific labeling method makes direct observation of polysaccharide gels challenging. Here, we exploit the specific interaction between $\kappa$-carrageenans and milk proteins (casein micelles), previously evidenced by electron microscopy and dynamic light scattering~\cite{Spagnuolo2005, Martin2006}, to track biopolymer assemblies in the samples, i.e., particles of fluid gels. In practice, $1~\%~(w/w)$ skim milk powder (low-heat, Ingredia, France) is added to the carrageenan solution, yielding a tracer volume fraction of approximately $1~\%~(v/v)$~\cite{Bauland2024a}. Milk proteins are stained with Fast Green FCF (Sigma-Aldrich) at a final concentration of $1~\rm{mg/L}$.

Fluid gels containing the protein tracers and obtained under various shear rates are prepared in the rheometer as previously described. A few milliliters of the sample are collected from the rheometer, gently transferred into four-well slides (\textmu-Slide 4 Well, Ibidi), and covered with mineral oil to prevent drying. Imaging is performed on the same day using a Yokogawa W1 spinning disk confocal microscope equipped with a $10\times$ objective ($0.45$ NA, CFI Plan Apo) and two EMCCD Andor iXon Ultra cameras ($1024\times1024$ pixels, $13\times13~\rm{\mu m}$ pixel size), yielding a spatial resolution of $1.29~\rm{\mu m}$. Fast Green FCF is excited with a 640~nm laser, and $z$-stacks of 120 images are acquired every $1.29~\rm{\mu m}$, corresponding to a sampled volume of $1321\times1321\times152~\rm{\mu m^3}$.

Image analysis is performed in MATLAB using functions from the Image Processing Toolbox. Background images are estimated by applying a Gaussian filter ($\sigma = 20$) with the \texttt{imgaussfilt3} function and subtracted from the raw images. Segmentation is achieved using hysteresis thresholding (\texttt{hysthreshold}). Briefly, structural features are first identified using a strong threshold, then extended by a weaker threshold to recover regions partially removed in the first step. In the resulting binary volumes, particles are identified using the \texttt{bwconncomp} function, and their volumes are measured with \texttt{regionprops3} to compute the equivalent spherical radius.

\begin{figure}[t!]
    \includegraphics[scale=0.5, clip=true, trim=0mm 0mm 0mm 0mm]{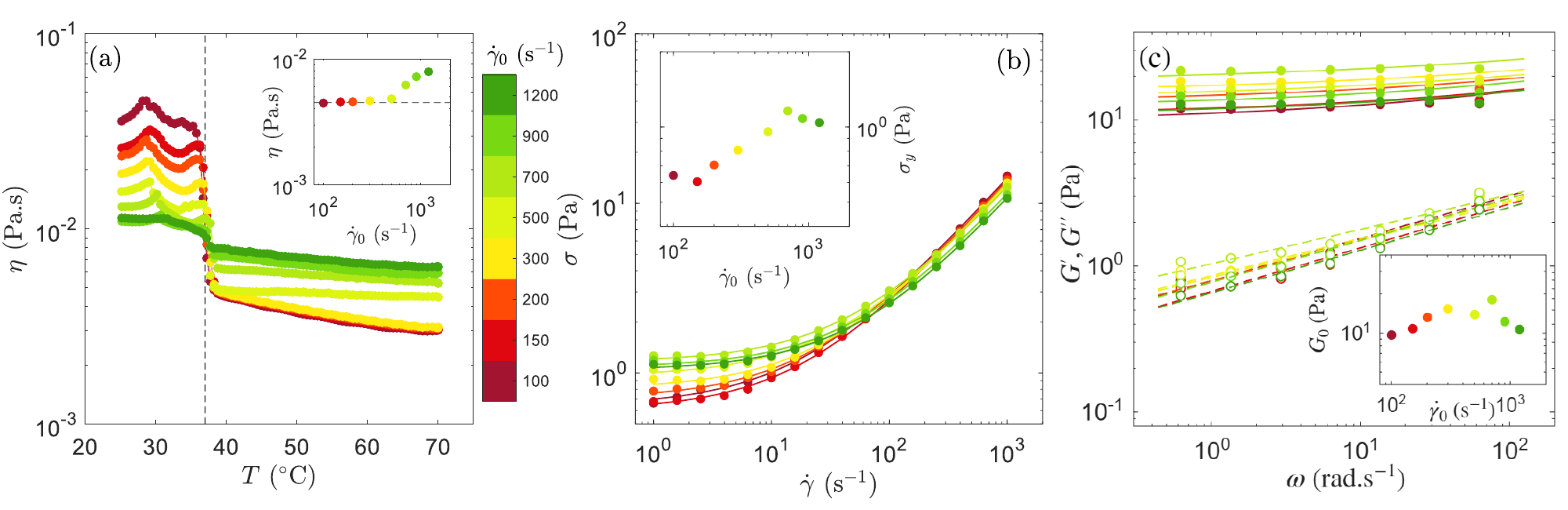}
    \centering
    \caption{
    Rheology of $\kappa$-carrageenan fluid gels with $C_w = 0.35~\%~(w/w)$ and $I = 40~\mathrm{mM}$. 
    (a) Apparent viscosity $\eta$ as a function of temperature $T$ during cooling under constant imposed shear rate $\dot{\gamma}_0$. The color scale indicates the applied shear rate. \jb{Vertical dotted line indicates the sol-gel transition at T = $37^{\circ}$C.} Inset: apparent viscosity measured at $70^{\circ}\mathrm{C}$ as a function of $\dot{\gamma}_0$. 
    (b) Flow curves of fluid gels formed at different shear rates, showing shear stress $\sigma$ versus $\dot{\gamma}$. The solid line represents the best fit with the Herschel–Bulkley model. Inset: dynamic yield stress $\sigma_y$ versus shear rate during cooling $\dot{\gamma}_0$. 
    (c) Viscoelastic spectra showing storage ($G^{\prime}$, filled symbols) and loss ($G^{\prime\prime}$, open symbols) moduli as a function of angular frequency $\omega$. Solid and dotted lines represent best fits with a fractional Kelvin–Voigt model. Inset: plateau modulus $G_0$ versus shear rate during cooling $\dot{\gamma}_0$.
    }
    \label{fig:shear}
\end{figure}

\section{Results and Discussion}

Typical rheometric experiments first involve heating the carrageenan solution to $70^{\circ}$C to erase any previous thermal or mechanical history, followed by cooling while applying a constant shear rate $\dot{\gamma}_0$, hereafter referred to as ``cooling shear rate''. In Fig.~\ref{fig:shear}(a), the viscosity measured at $70^{\circ}$C for $0.35~\%~(w/w)$ $\kappa$-carrageenan with $40~\rm{mM}$ KCl is roughly constant, about $4.5~\rm{mPa \, s}$, and sharply increases at $T \approx 37^{\circ}$C, marking the sol–gel transition. The inset of Fig.~\ref{fig:shear}(a) shows that the sol-phase viscosity at high temperature is Newtonian. The deviation observed at high shear rates ($\dot{\gamma}_0 \geq 700~\rm{s}^{-1}$) arises from inertial effects and hydrodynamic instabilities which lead to enhanced dissipation and an increased apparent viscosity (see Section~`Taylor–Couette Instability' in Appendix). 

After cooling to $25^{\circ}$C, flow properties of the resulting fluid gels are assessed by ramping the shear rate down from $1000$ to $1~\rm{s}^{-1}$ while measuring the shear stress $\sigma$ [Fig.~\ref{fig:shear}(b)]. Applying such high shear rates ensures that differences between samples originate from the shear rate imposed during cooling ($\dot{\gamma}_0$), rather than from the maximum shear rate experienced. The flow curves are fitted with the Herschel–Bulkley model, $\sigma = \sigma_y[1 + (\dot{\gamma}/\dot{\gamma}_c)^{n}]$, allowing extraction of the Herschel–Bulkley or dynamic yield stress $\sigma_y$~\cite{Dinkgreve2016}. \jb{The other parameters, a characteristic shear rate $\dot{\gamma}_c$ and a dimensionless exponent $n$, characterize the onset and the magnitude of shear-thinning behavior ($n<1$), respectively.} The inset of Fig.~\ref{fig:shear}(b) reports $\sigma_y$ as a function of $\dot{\gamma}_0$. Initially, $\sigma_y$ increases with $\dot{\gamma}_0$, from about $0.03$ up to $0.3~\rm{Pa}$, after which it plateaus for $\dot{\gamma}_0 > 500~\rm{s}^{-1}$.

After flow cessation, the fluid gels are left to age for $1000~\rm{s}$ (see Fig.~S1 in the Appendix) before measuring their viscoelastic spectra, i.e., $G^{\prime}$ and $G^{\prime\prime}$ as a function of $\omega$, using a frequency sweep [Fig.~\ref{fig:shear}(c)]. The spectra display a gel-like response, characterized by an essentially frequency-independent storage modulus $G^{\prime}$, and this overall shape is insensitive to the shear rate applied during cooling. The inset of Fig.~\ref{fig:shear}(c) shows the corresponding plateau modulus $G_0$ as a function of $\dot{\gamma}_0$, revealing no systematic dependence of the gel elasticity on the cooling shear rate, with $G_0 \approx 10~\rm{Pa}$.

\begin{figure}[t!]
    \includegraphics[scale=0.53, clip=true, trim=0mm 0mm 0mm 0mm]{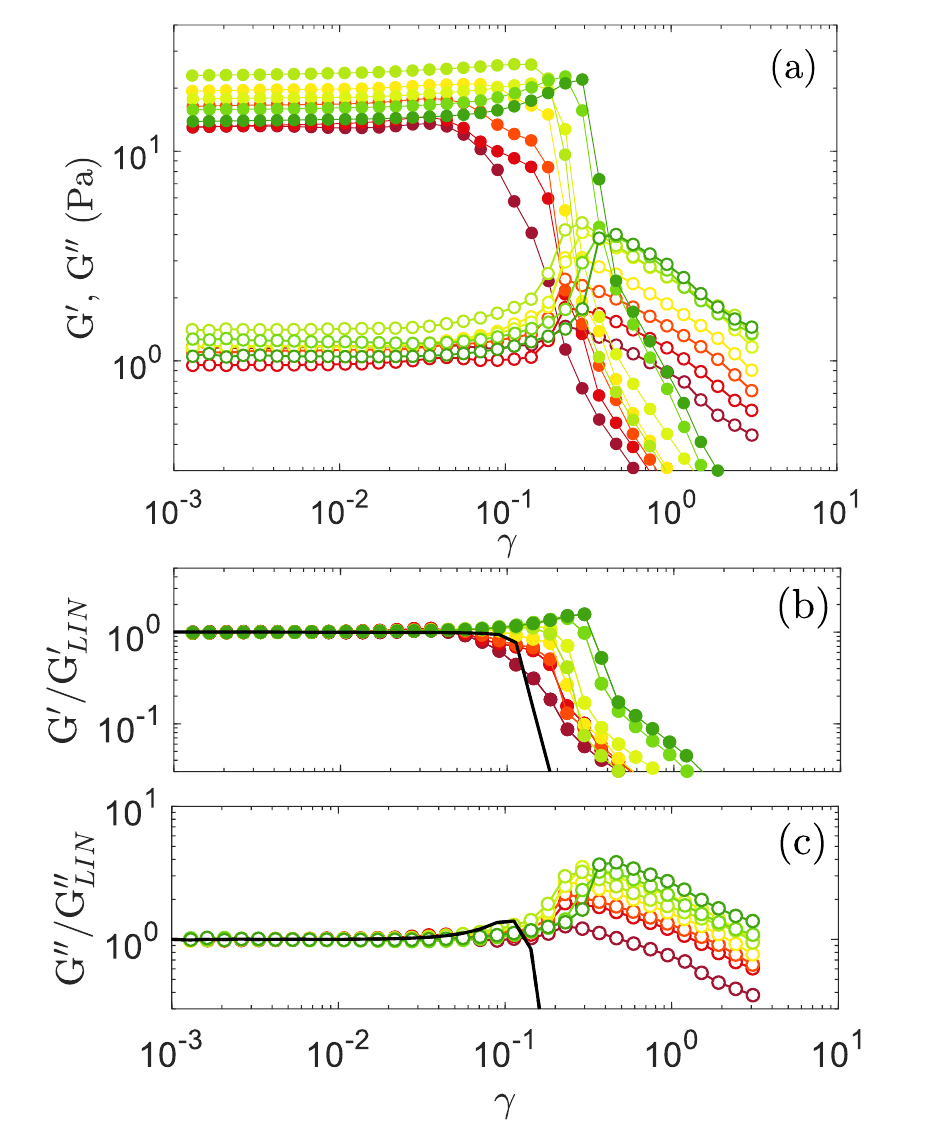}
    \centering
    \caption{
    Nonlinear viscoelasticity of $\kappa$-carrageenan fluid gels with $C_w = 0.35~\%~(w/w)$  and $I = 40~\mathrm{mM}$ measured from strain amplitude sweeps. 
    (a) Storage modulus $G^{\prime}$ (filled symbols) and loss modulus $G^{\prime\prime}$ (open symbols) as a function of shear strain $\gamma$. The color scale indicates the shear rate applied during cooling, as in Fig.~\ref{fig:shear}. 
    (b,c) $G^{\prime}$ and $G^{\prime\prime}$ normalized by their linear-regime values, $G^{\prime}_{\mathrm{LIN}}$ and $G^{\prime\prime}_{\mathrm{LIN}}$, as a function of $\gamma$. The solid black line shows the response of the gel formed under quiescent conditions ($\dot{\gamma}_0 = 0~\mathrm{s^{-1}}$).
    }

    \label{fig:sweep}
\end{figure}

We now examine the effect of shear applied during cooling on the nonlinear viscoelastic response, assessed by strain amplitude sweeps. Fig.~\ref{fig:sweep}(a) displays $G^{\prime}$ and $G^{\prime\prime}$ as a function of the shear strain $\gamma$. At low strain amplitudes, $\gamma < 5 \times 10^{-2}$, the moduli are constant, as expected in the linear regime, with $G^{\prime} > G^{\prime\prime}$. When the strain reaches the yield strain $\gamma_y$, $G^{\prime}$ drops sharply, signaling the  sudden loss of the load-bearing network, until $G^{\prime} < G^{\prime\prime}$ and complete fluidization. 
During this transition, $G^{\prime\prime}$ exhibits an overshoot, a characteristic hallmark of plastic flow in soft solids and gels~\cite{Donley2020}. 

In Fig.~\ref{fig:sweep}(b,c), $G^{\prime}$ and $G^{\prime\prime}$ are normalized by their linear-regime values to facilitate comparison across cooling shear rates $\dot{\gamma}_0$. 
For low $\dot{\gamma}_0$, the yield strain $\gamma_y$ is small, and the $G^{\prime\prime}$ overshoot remains weak, similar to the gel formed under quiescent conditions [solid curve in Fig.~\ref{fig:sweep}(b,c)]. For $I$= 40 mM, the gel formed under quiescent conditions is thus more brittle that the gels formed under shear. In contrast, for $I = 20$ and 30~mM, the gels formed at rest are more ductile than the gels formed under shear (see Fig.~S2 in the Appendix). For the latter, as $\dot{\gamma}_0$ increases, both $\gamma_y$ and the amplitude of the $G^{\prime\prime}$ overshoot increase [Fig.~\ref{fig:sweep}(c)]. Accordingly, the non-linear properties of fluid gels evolve continuously with the cooling shear rate.

\begin{figure}[t!]
    \includegraphics[scale=0.53, clip=true, trim=0mm 0mm 0mm 0mm]{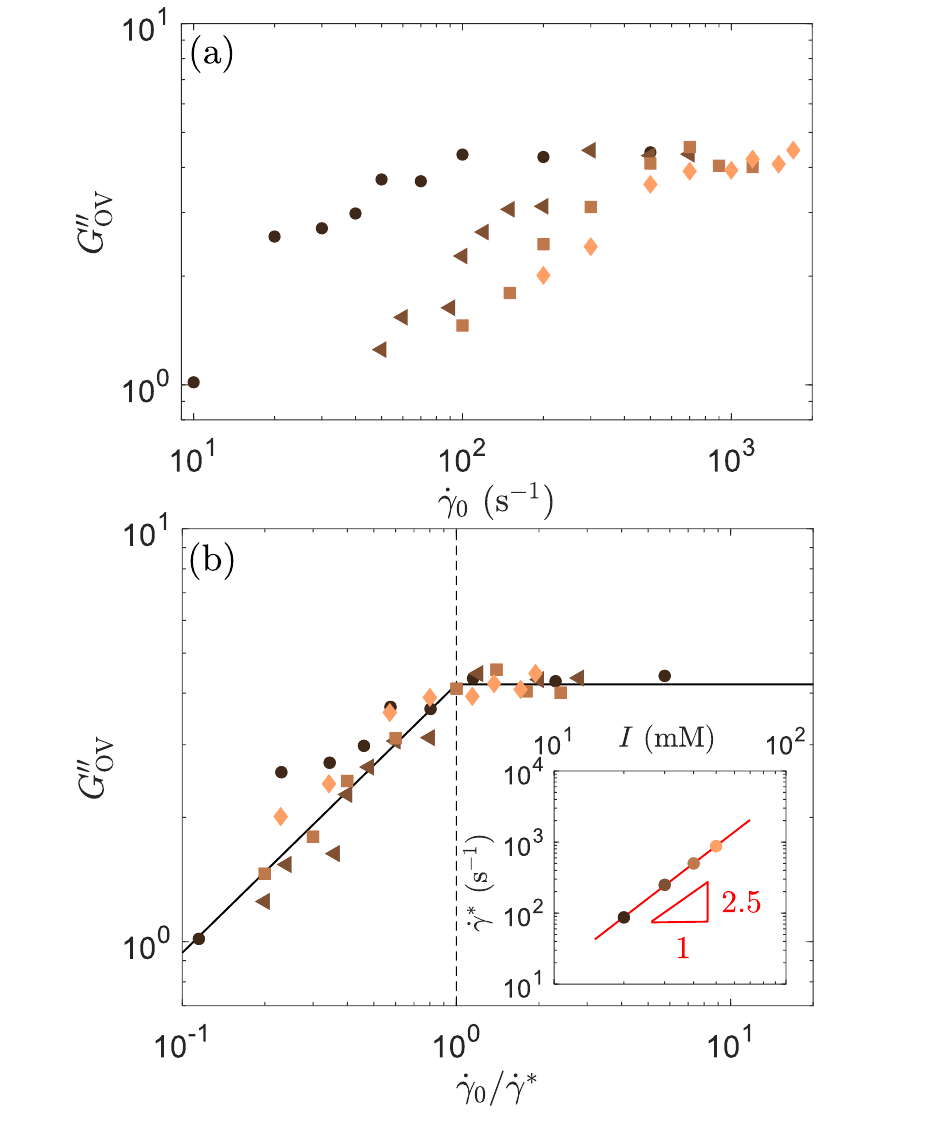}
    \centering
    \caption{
    (a) Magnitude of the overshoot in $G^{\prime\prime}$ measured during strain amplitude sweeps as a function of the preshear rate $\dot{\gamma}_0$ applied during cooling. Marker shape indicates the KCl-adjusted ionic strength: $I = 20$~mM (circles), 30~mM (triangles), 40~mM (squares), and 50~mM (diamonds). 
    (b) $G^{\prime\prime}$ overshoot magnitude as a function of the normalized preshear rate $\dot{\gamma}_0 / \dot{\gamma}^*$. Inset: characteristic shear rate $\dot{\gamma}^*$ versus ionic strength $I$. The red line represents the best power-law fit with an exponent of 2.5.
    }
    \label{fig:mastercurve}
\end{figure}

To assess the generality of this behavior, we vary the ionic strength $I$ of the carrageenan solutions between $20$ and $50~\rm{mM}$ and repeat the same protocol (see Fig.~S1 and Fig.~S2 in the Appendix). We use the normalized amplitude of the $G^{\prime\prime}$ overshoot, $G^{\prime\prime}/G^{\prime\prime}_{\mathrm{LIN}}$, as a marker of the non-linear response. Its evolution with $\dot{\gamma}_0$ is shown in Fig.~\ref{fig:mastercurve}(a) for different $I$. For all ionic strengths tested, the overshoot amplitude increases with $\dot{\gamma}_0$ and plateaus beyond an upper shear rate, which itself depends on $I$. This indicates that the non-linear properties become insensitive to further increases in $\dot{\gamma}_0$ once this threshold is reached. Such behavior is consistent with the evolution of the dynamic yield stress with $\dot{\gamma}_0$, observed for $I = 40~\rm{mM}$ [inset of Fig.~\ref{fig:shear}(b)]. In Fig.~\ref{fig:mastercurve}(b), all curves collapse onto a single master curve when the shear rate is rescaled by a critical value $\dot{\gamma}^*$, which marks the onset of the plateau in the $G^{\prime\prime}$ overshoot. This collapse indicates that, irrespective of ionic strength, the effect of shear on the non-linear properties follows the same phenomenology and is governed by a distinct critical shear rate. As shown in the inset of Fig.~\ref{fig:mastercurve}(b), $\dot{\gamma}^*$ scales with ionic strength as $\dot{\gamma}^* \propto I^{2.5}$.

\begin{figure}[t!]
    \includegraphics[scale=0.53, clip=true, trim=0mm 0mm 0mm 0mm]{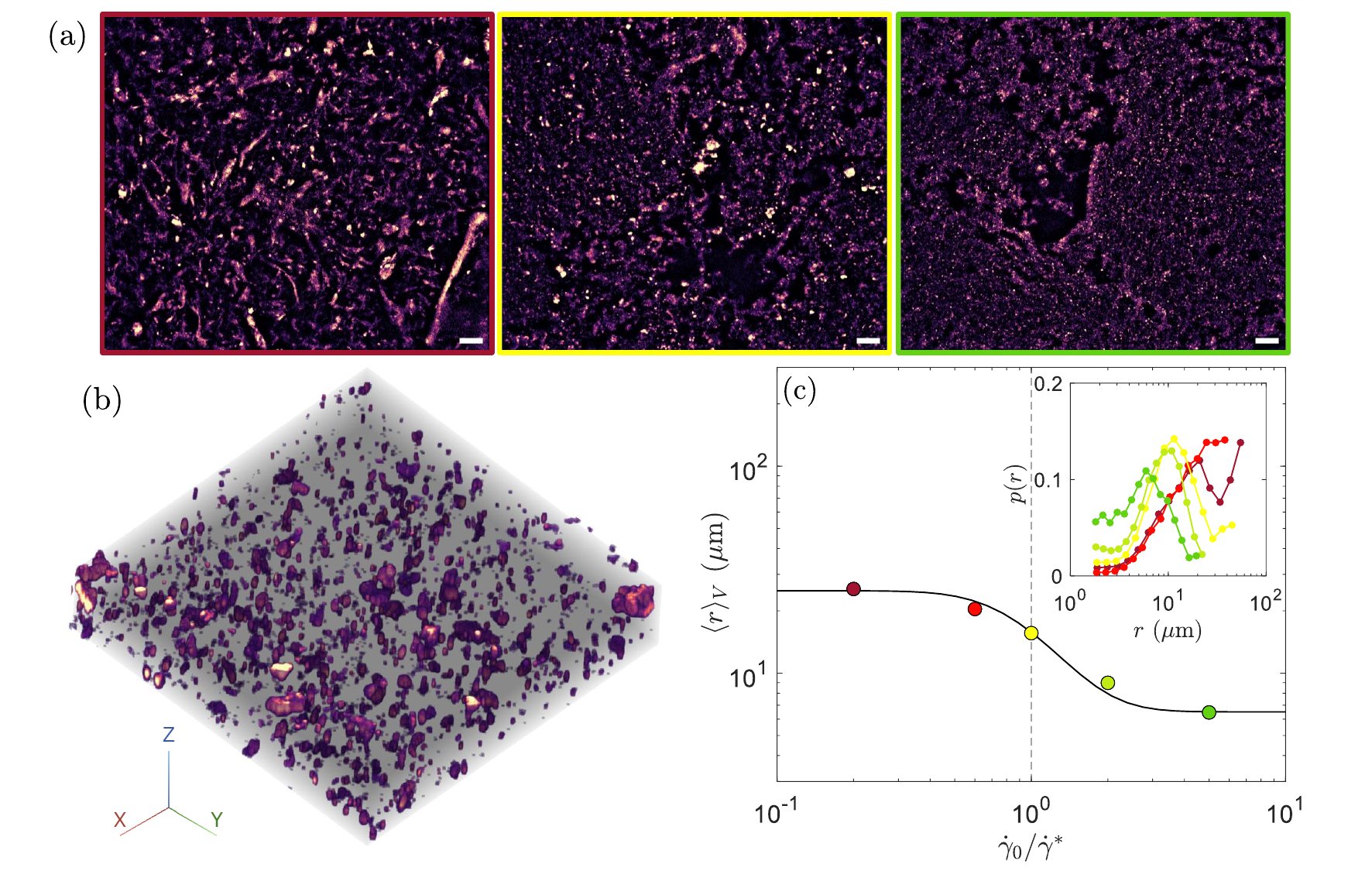}
    \centering
    \caption{
    Confocal microscopy images of $\kappa$-carrageenan fluid gels formed at different shear rates during cooling. 
    (a) Representative images of gels formed at (from left to right) $\dot{\gamma}_0 = \dot{\gamma}^*/5$, $\dot{\gamma}^*$, and $5 \times \dot{\gamma}^*$. [Scale bar: $100~\mu\mathrm{m}$]. 
    (b) Example of a 3D volume reconstruction for the gel formed at $\dot{\gamma}_0 = \dot{\gamma}^*$.
    (c) Volume-weighted average particle radius as a function of the rescaled shear rate $\dot{\gamma}_0 / \dot{\gamma}^*$. Inset: probability distribution function of particle size.
    }
    \label{fig:micro}
\end{figure}

To gain insights into the microscopic mechanisms underlying this two-step process, separated by $\dot{\gamma}^*$, we probe the structure of fluid gels with $I \approx 30~\rm{mM}$ using confocal microscopy (see Materials and Methods Section and Fig.~S5 in the Appendix for experimental details). Fig.~\ref{fig:micro}(a) shows images of fluid gels formed under cooling shear rates of $\dot{\gamma}_0 = \dot{\gamma}^*/5$, $\dot{\gamma}^*$, and $5 \times \dot{\gamma}^*$ (left to right), where polymeric particles appear in warm colors. All samples exhibit heterogeneous microstructures composed of micrometric particles, in agreement with earlier studies~\cite{Garrec2012,Gabriele2009,DOria2023a}. Visual inspection indicates that particle size decreases with increasing $\dot{\gamma}_0$. Quantitative analysis of particle size is performed from the 3D volume reconstruction of the acquired $z$-stacks, as illustrated in Fig.~\ref{fig:micro}(b). Fig.~\ref{fig:micro}(c) reports the volume-weighted average particle radius $\langle r \rangle_V$ as a function of the rescaled cooling shear rate $\dot{\gamma}_0 / \dot{\gamma}^*$. With increasing shear, the particle radius decreases from roughly $25~\mu\mathrm{m}$ to about $6~\mu\mathrm{m}$, with a sharp transition occurring at $\dot{\gamma}_0 = \dot{\gamma}^*$. Thus, the transition in the non-linear properties with $\dot{\gamma}_0$ originates from the flow-induced reduction in particle size, which decreases with increasing $\dot{\gamma}_0$ and remains essentially constant for $\dot{\gamma}_0 > \dot{\gamma}^*$. In addition to a reduction of the aggregate size, previous studies on shear-induced structuring of particulate gels have reported a densification of the aggregates with increasing shear rate~\cite{Masschaele2011,Colombo2025}. Here, our structural measurements, which prioritize large fields of view and statistical robustness for particle size determination, do not allow us to probe the internal mass distribution within the particles. Nonetheless, the particle size obtained in the high-shear limit, $r_V \approx 6.5~\mu\mathrm{m}$, is in excellent agreement with the value of $5~\mu\mathrm{m}$ previously reported for $\kappa$-carrageenan fluid gels in the presence of KCl~\cite{Gabriele2009}. \jb{Note that heterogeneous flow fields have been reported during the settling under shear of 2\% (w/w) $\kappa$-carrageenan dispersions at low shear rates ($\dot{\gamma} = 1~\rm{s}^{-1}$)~\cite{Hilliou2014}. In the present study, where we rely on higher rates, potential shear localization and the resulting effect on the microstructure would lead to an effective particle size, as the material is re-dispersed at high rates during the flow curves.}

In conclusion, the thermo-rheological memory of carrageenan fluid gels, as manifested in their non-linear properties, results from the competition between shear magnitude and particle size, governed by a critical shear rate $\dot{\gamma}^*$. When $\dot{\gamma}_0 < \dot{\gamma}^*$, shear forces are weak compared to the attractive interactions between carrageenan chains, allowing the formation of large structures that can survive the flow field. The lower limit corresponds to a gel formed at rest, i.e., $\dot{\gamma}_0 = 0~\rm{s}^{-1}$, which is subsequently fluidized into a particulate dispersion when performing the flow curve starting at $1000~\rm{s}^{-1}$. When $\dot{\gamma}_0 > \dot{\gamma}^*$, shear forces become strong enough to impose, through aggregation–disaggregation processes, a maximum particle size that can be sustained under flow.

For dispersions of attractive particles, structuring under flow is commonly rationalized as a competition between the viscous drag force acting on particles ($F_\mathrm{visc}$) and the attractive force between them ($F_\mathrm{attr}$), captured by the Mason number~\cite{Mewis2009,Jamali2020}:

\begin{equation}
Mn = \frac{F_\mathrm{visc}}{F_\mathrm{attr}} = \frac{6 \eta_f a^2 \dot{\gamma}}{U / \delta},
\label{eq:Mn}
\end{equation}

where $\eta_f$ is the viscosity of the continuous phase \jb{(here water)}, and $U$ and $\delta$ are the depth and range of the attractive potential between particles of size $a$.

Note that for Brownian particles, the relevant interaction strength must account for thermal activation of the bonds, and is appropriately defined as the most probable rupture force $f^*$ with $f^* < U/\delta$.~\cite{Varga2018,Colombo2025}
Values of $Mn > 1$ indicate that hydrodynamic forces dominate, while $Mn < 1$ indicates that attractive forces prevail, consistent with the microstructural scenario revealed by our data: both rheology and microscopy suggest that $Mn \approx 1$ when $\dot{\gamma} = \dot{\gamma}^*$.

As shown in Fig.~\ref{fig:mastercurve}(b), the critical shear rate depends on the adjusted ionic strength, following $\dot{\gamma}^* \propto I^{2.5}$. We aim to rationalize this scaling by first turning to the rheology of gels formed under quiescent conditions (see Fig.~S3 in the Appendix). We find that the gel elasticity, assessed from the plateau modulus $G_0$, increases as $G_0 \propto I^{2.5}$ for $I \in [10\text{–}50]~\rm{mM}$, and then appears to plateau for $I > 50~\rm{mM}$. This macroscopic evolution closely mirrors the ionic-strength dependence of the critical shear rate and reflects the ion-mediated coil–helix transition characteristic of $\kappa$-carrageenan gels~\cite{Schefer2015a,Schefer2015}. As the ion concentration increases, flexible coils progressively self-assemble into bundles of rigid helices, leading to an increase in the effective persistence length of the semi-flexible chains. In particular, upon addition of KCl, the coil–helix transition is reported to begin above roughly $7~\rm{mM}$ and to be complete near $50~\rm{mM}$~\cite{Schefer2015}, in good agreement with the increase of $G_0$ observed for quiescent gels in our experiments. This correspondence points to a direct link between the molecular stiffening induced by the coil–helix transition and the macroscopic shear scale governing gel formation under flow.

Similarly to the Mason number, defined as a ratio between forces, other related dimensionless groups can be constructed in the form of an Adhesion number ($Ad$)~\cite{Bouthier2023}:

\begin{equation}
\label{eq:ad}
    Ad = (Mn)^{-1} = \frac{U}{\sigma_f \delta^{3}}
\end{equation}

where $\sigma_f = \dot{\gamma}\eta_f$ is the viscous stress.

Written in this form, $U/\delta^3$ is an energy per unit volume, expressed in Pascals like $\sigma_f$, so that their ratio is indeed dimensionless. For a homogeneous network with a well-defined mesh size, such as gels formed under quiescent conditions, the shear modulus $G_0$, also expressed in Pascals, is expected to scale with the elastic energy density of the system, yielding $G_0 \sim U/\delta^3$. By dimensional analysis and from Eq.~\ref{eq:ad}, the Adhesion number can therefore be written as $Ad = G_0/\sigma_f$. For $Ad = 1$, with $\sigma_f = \dot{\gamma}^*\eta_f$, the critical shear rate scales as $\dot{\gamma}^* \propto G_0 \propto I^{2.5}$. This rationalizes the identical scalings of $\dot{\gamma}^*$ under flow and the plateau modulus $G_0$ of quiescent gels with ionic strength. The Adhesion number thus compares the elastic energy density of the system ($G_0$), shaped by ion-specific self-assembly of $\kappa$-carrageenan chains, with the viscous stress ($\sigma_f$).

In its most general form, the Adhesion number can be expressed as
$Ad = U / (\sigma_f a^{\alpha} \delta^{3-\alpha})$, with $\alpha \in [0,3]$ a dimensionless parameter that depends on whether the ratio is constructed from energies ($\alpha = 0$), forces ($\alpha = 1$), or stiffnesses ($\alpha = 2$), corresponding respectively to energies per unit surface, or energies per unit volume ($\alpha = 3$)~\cite{Bouthier2023}. Choosing $\alpha = 3$ is a natural choice when the bulk elasticity of the particle or gelling network controls aggregation and breakup, as predicted by affine network models widely used to describe semi-flexible fiber networks such as actin~\cite{Mackintosh1995,Broedersz2014}. These models assume affine deformation of the network under load, directly linking the bulk elastic modulus to the internal energy density $U/\delta^3$.

From its definition in Eq.~\ref{eq:ad}, the Adhesion number is insensitive to the polymer concentration. \jb{Indeed, for dilute suspensions, we consider the ratio of an energy density, defined over the volume of a single fluid gel particle, to the viscous stress of the solvent, which is consistent with the dilute regime considered here.} Consequently, the critical shear rate $\dot{\gamma}^*$ at which $Ad=1$, is expected to be independent of biopolymer concentration, provided that the viscosity of the sol-phase (at high temperature) remains roughly constant. Fig.~S4 in the Appendix confirms this expectation: reducing or increasing carrageenan concentration by a factor 2 does not alter $\dot{\gamma}^*$, supporting the idea that fluid gel structuring under flow is governed by an Adhesion number. 

Note that the non-linear properties, assessed by the amplitude of the $G^{\prime\prime}$ overshoot (Fig.~\ref{fig:mastercurve}), and the particle size [Fig.~\ref{fig:micro}(c)] both plateau in the high-shear-rate limit, while the Adhesion number continues to decrease. This plateau behavior may be attributed to the emergence of inertial effects for $\dot{\gamma} \geq 700~\rm{s}^{-1}$. This hypothesis could be tested by increasing the solvent viscosity, which would (i) reduce inertial effects over the investigated range of shear rates and (ii) increase the viscous drag force $\sigma$ acting on the particles. This will be investigated in future work.

\section{Conclusion}
In conclusion, we have shown that $\kappa$-carrageenan fluid gels formed under flow encode the shear history of gelation mainly in their nonlinear viscoelastic response, while their linear elasticity remains comparatively insensitive to the cooling shear rate. Strain-amplitude sweeps reveal a robust two-regime behavior: the magnitude of the $G^{\prime\prime}$ overshoot increases with the imposed cooling shear rate $\dot{\gamma}_0$ up to a critical value $\dot{\gamma}^*$, beyond which it plateaus. Confocal microscopy directly links this rheological memory to a shear-selected particle size, with a sharp reduction in the characteristic radius occurring at the same $\dot{\gamma}^*$. \jb{This memory is encoded in the microstructure by the shear rate imposed during the thermally controlled sol-gel transition, irrespective of the maximum shear rate experienced afterwards, and is hence termed ``thermo-rheological memory.''} By varying ionic strength, we further demonstrate that $\dot{\gamma}^*$ scales as $\dot{\gamma}^* \propto I^{2.5}$, mirroring the scaling of the plateau modulus of quiescent gels and implicating the ion-mediated coil--helix transition as the molecular origin of this dependence. These observations are unified by an Adhesion number, $Ad \sim G_0/(\eta_f \dot{\gamma})$, which compares the elastic energy density, set by chain self-assembly, to the viscous stress imposed by flow; the condition $Ad \approx 1$ naturally defines $\dot{\gamma}^*$ and rationalizes its ionic-strength scaling. The present results are reminiscent of shear-induced structuring and flow-history-dependent rheology in particulate gel dispersions~\cite{Masschaele2011,Koumakis2015,Sudreau2022a,Varga2018}, thus bridging the gap between colloidal gels and gels made of macromolecules. Beyond providing a quantitative framework for thermo-rheological memory in fluid gels, this work establishes a processing-based route to tune microstructure and dissipation without changing composition, with direct implications for formulation stability, yielding, and performance under complex flows.

\begin{acknowledgement}

Authors acknowledge Nestlé Research (Lausanne, Switzerland) for financial support. Authors acknowledge the help of Tobias Schwarz and Joachim Hehl (ScopeM, ETH Zürich) for their help the the confocal microscope.  
\end{acknowledgement}

\begin{suppinfo}
Supporting Information is available and includes: (i) rheological data for fluid gels at all ionic strengths; (ii) rheological data for quiescent gels; (iii) rheological characterization in the presence of tracers for microscopy imaging; and (iv) a discussion of the Taylor-Couette instability.
\end{suppinfo}

\bibliography{library}

\end{document}


\maketitle

\section{Additional Rheology}

\subsection{Data at various Ionic Strength}

\begin{figure}[t!]
    \includegraphics[scale=0.6, clip=true, trim=0mm 0mm 0mm 0mm]{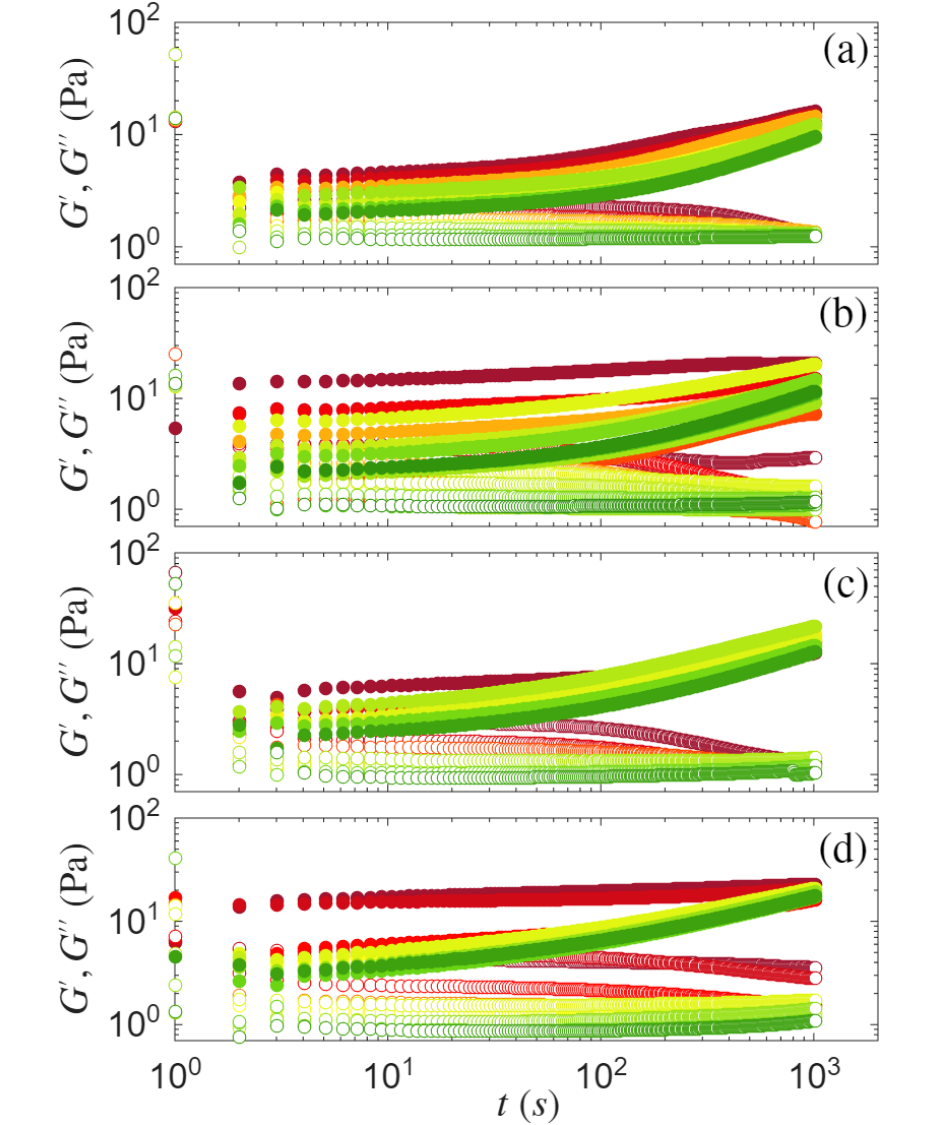}
    \centering
    \caption{Aging of $\kappa$-carrageenan fluid gels with $C_w = 0.35~\%~(w/w)$ after being brought to rest, assessed by the temporal evolution of the storage ($G^{\prime}$, filled symbols) and loss ($G^{\prime\prime}$, open symbols) moduli. Panels (a–d) correspond to ionic strengths $I = 20$, $30$, $40$, and $50~\mathrm{mM}$, respectively. The color scale indicates the shear rate applied during cooling, as in Fig.~1.
    }
    \label{figsupp:aging}
\end{figure}

Fig.~\ref{figsupp:aging} shows the aging kinetics of $0.35~\%~(w/w)$ fluid gels at various ionic strengths after being brought to rest. For all ionic strengths and cooling shear rates, fluid gels exhibit aging, i.e., an increase of $G^{\prime}$ over time, while $G^{\prime\prime}$ remains roughly constant. Only gels formed at the lowest shear rates (dark red markers) show weaker aging, with $G^{\prime}$ remaining nearly constant over time. The similarity of the aging kinetics indicates that the trend of $G_0$ versus $\dot{\gamma}_0$ depicted in Fig.~1(c) in the main text does not depend on the aging time of the samples, here fixed at $t = 1000~\rm{s}$.

\begin{figure*}[t!]
    \includegraphics[scale=0.6, clip=true, trim=0mm 0mm 0mm 0mm]{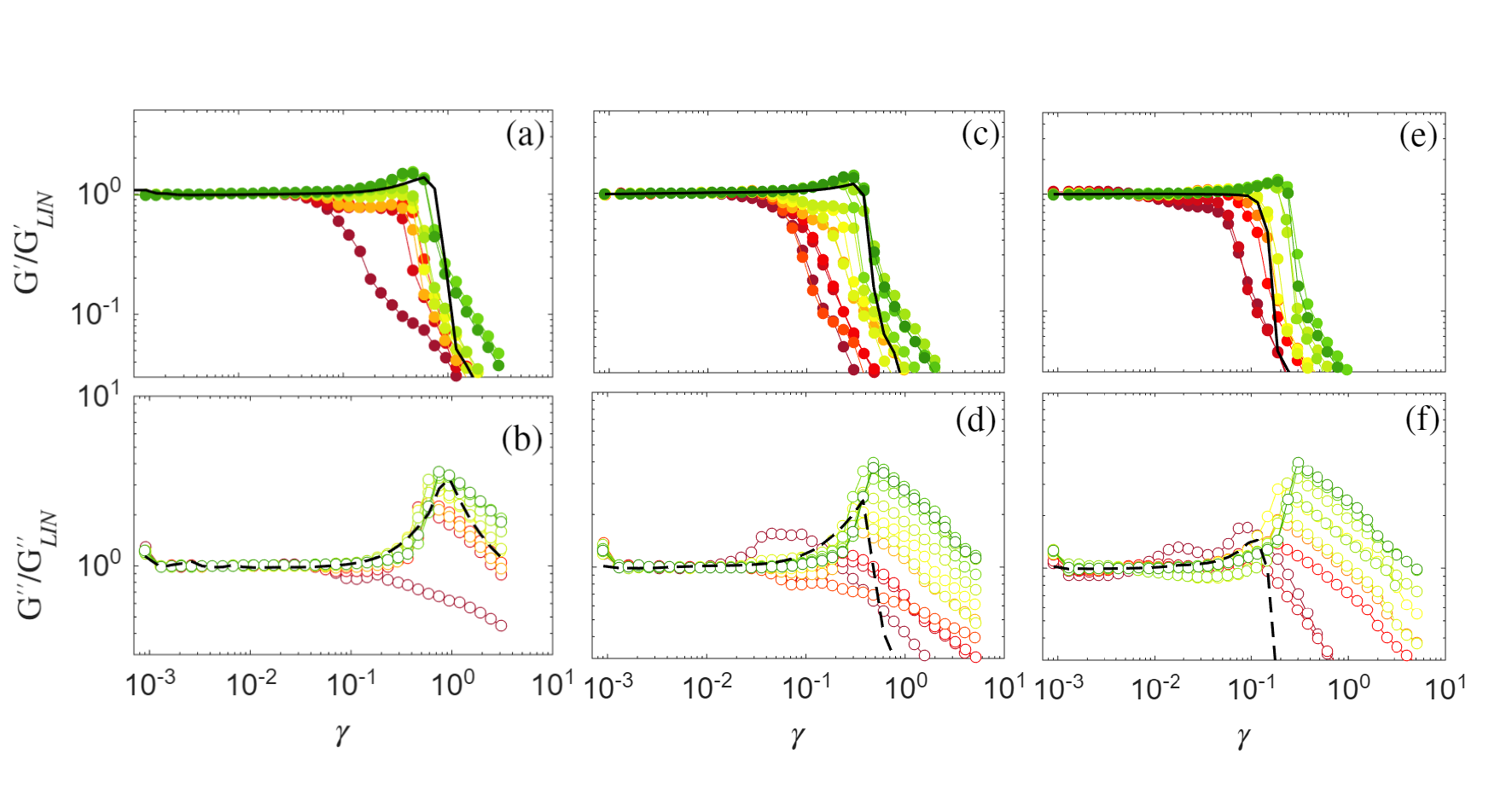}
    \centering
    \caption{Nonlinear viscoelasticity of $\kappa$-carrageenan fluid gels with $C_w = 0.35~\%~(w/w)$ at (a,b) $I = 20~\mathrm{mM}$, (c,d) $I = 30~\mathrm{mM}$, and (e,f) $I = 50~\mathrm{mM}$, measured from strain-amplitude sweeps. The storage modulus $G^{\prime}$ (filled symbols) and loss modulus $G^{\prime\prime}$ (open symbols) are normalized by their linear-regime values, $G^{\prime}{\mathrm{LIN}}$ and $G^{\prime\prime}{\mathrm{LIN}}$, and plotted as a function of the shear strain $\gamma$. The color scale indicates the shear rate applied during cooling, as in Fig.~1. The solid black line shows the response of a gel of the same composition formed under quiescent conditions ($\dot{\gamma}_0 = 0~\mathrm{s^{-1}}$).
    }
    \label{figsupp:sweep}
\end{figure*}

Fig.~\ref{figsupp:sweep} shows the strain-amplitude sweeps of $0.35~\%~(w/w)$ fluid gels at various ionic strengths. Gels of identical composition but formed under quiescent conditions are shown as black lines. Under significant load, ductile materials are able to dissipate energy through plastic activity before yielding, in contrast with brittle materials for which yielding is abrupt and catastrophic~\cite{Divoux2024}. The $G^{\prime\prime}$ overshoot is a characteristic hallmark of such plastic flow in soft solids and gels, preceding failure~\cite{Donley2020}. As depicted in Fig.~\ref{figsupp:sweep}, the magnitude of the $G^{\prime\prime}$ overshoot for gels formed at rest strongly depends on ionic strength, being pronounced at $I = 20~\mathrm{mM}$ and almost nonexistent at $I = 50~\mathrm{mM}$ [Fig.~\ref{figsupp:sweep}(b) and (f), respectively]. In contrast, the $G^{\prime\prime}$ overshoot of fluid gels is robust with respect to ionic strength and depends primarily on the critical shear rate $\dot{\gamma}^*$, as explained in the main text. As a result, ductile materials can be obtained at any ionic strength, with a $G^{\prime\prime}$ overshoot peaking at $\gamma \approx 1$, provided that the appropriate shear rate is applied during cooling.

\subsection{Quiescent Gels}

\begin{figure}[t!]
    \includegraphics[scale=0.6, clip=true, trim=0mm 0mm 0mm 0mm]{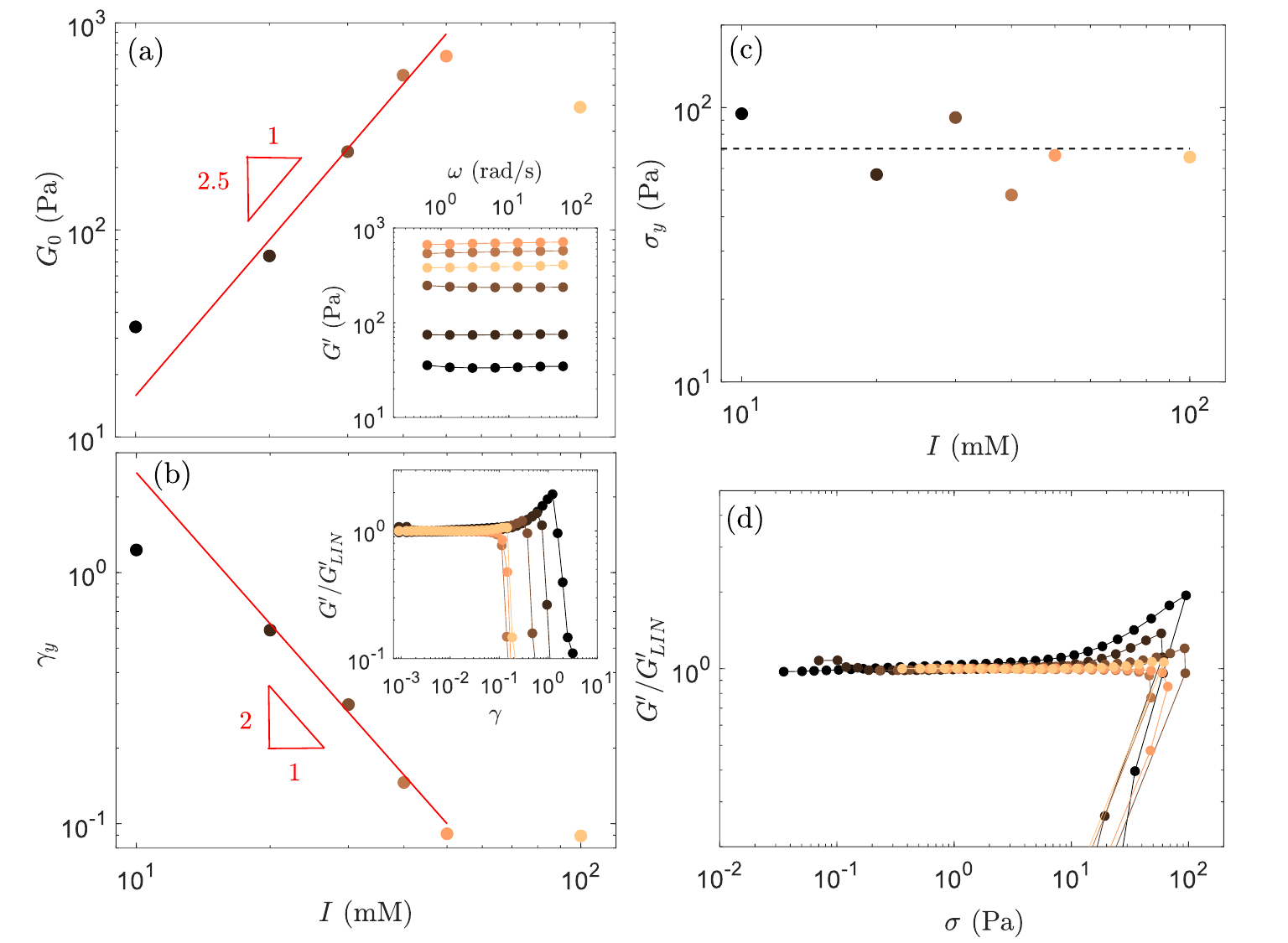}
    \centering
    \caption{Rheology of $C_w = 0.35~\%~(w/w)$  $\kappa$-carrageenans gels formed under quiescent conditions ($\dot{\gamma}_0 = 0~\rm{s}^{-1}$). (a) Elasticity (plateau modulus) $G_0$ versus ionic strength $I$. Red curve is the best power law fit, yielding a power exponent 2.5. Inset: storage modulus $G^{\prime}$ versus angular frequency $\omega$ for the corresponding gels. (b) Strain at failure $\gamma_y$ versus ionic strength $I$. Red curve is the best power law fit, yielding a power exponent -2. Inset: normalized storage modulus $G^{\prime}/G^{\prime}_{LIN}$ versus strain $\gamma$ for the corresponding gels. (c) Stress at failure $\sigma_y$ versus ionic strength $I$. Black dotted line is the averaged $\sigma_y$ value for all $I$. (d) Normalized storage modulus $G^{\prime}/G^{\prime}_{LIN}$ versus stress $\gamma$.
    }
    \label{figsupp:gelrest}
\end{figure}

Fig.~\ref{figsupp:gelrest}(a) shows the plateau modulus $G_0$ of gels formed under quiescent conditions at various ionic strengths $I$. $G_0$ increases as $I^{2.5}$ up to 50~mM, after which it appears to plateau. As discussed in the main text, this macroscopic trend is fully consistent with the ion-induced transition of carrageenan chains from flexible coils to rigid helices. Atomic Force Microscopy (AFM) imaging has shown that the coil–helix transition of $\kappa$-carrageenans in the presence of KCl starts as low as 0.1~mM, while strands or bundles of helices appear from $I = 7$~mM. The strand configuration dominates at 50~mM~\cite{Schefer2015}. Further salt addition is expected to reduce chain repulsions, but as not additional effect on the chains microstructure. This microscopic evolution aligns with the increase in gel elasticity, supporting a direct link to chain assembly.

The elasticity of cross-linked semi-flexible polymers, including biopolymers such as actin, can be predicted using affine network models derived from non-classical rubber theory. The network elasticity depends on the bending modulus $k$ of the chains, which itself depends on the persistence length $l_p$, $k = K T l_p$, and on the network mesh size $\xi$. In the densely cross-linked regime, the storage modulus scales as:

\begin{equation}
    G^{\prime} \sim \frac{k^2}{K T} \xi^{-5} \sim \frac{k^2}{K T} (a c_A)^{5/2}
\end{equation}

where $K$ is the Boltzmann constant, $T$ the temperature, $a$ the chain monomer size, and $c_A$ the chain concentration. As shown in the inset of Fig.~\ref{figsupp:conc}(c), $G^{\prime}$ scales with concentration as $G^{\prime} \sim c_A^{2.4}$, in agreement with affine network predictions. A complete modeling of carrageenan gel elasticity, however, is beyond the scope of this work. Nonetheless, it is interesting to note that the elastic modulus of fluid gels is between 1 and 2 orders of magnitude lower, depending on the ionic strength, as compared to gels formed under quiescent conditions. While elasticity in the quiescent gel arises from a continuous network of bundled fibers~\cite{Schefer2015a}, captured by an effective persistence length, fluid gels possess a particulate microstructure involving fewer contact points and weak, lateral interactions between already structured chains.

\subsection{Carrageenans Concentration}

\begin{figure}[t!]
    \includegraphics[scale=0.53, clip=true, trim=0mm 0mm 0mm 0mm]{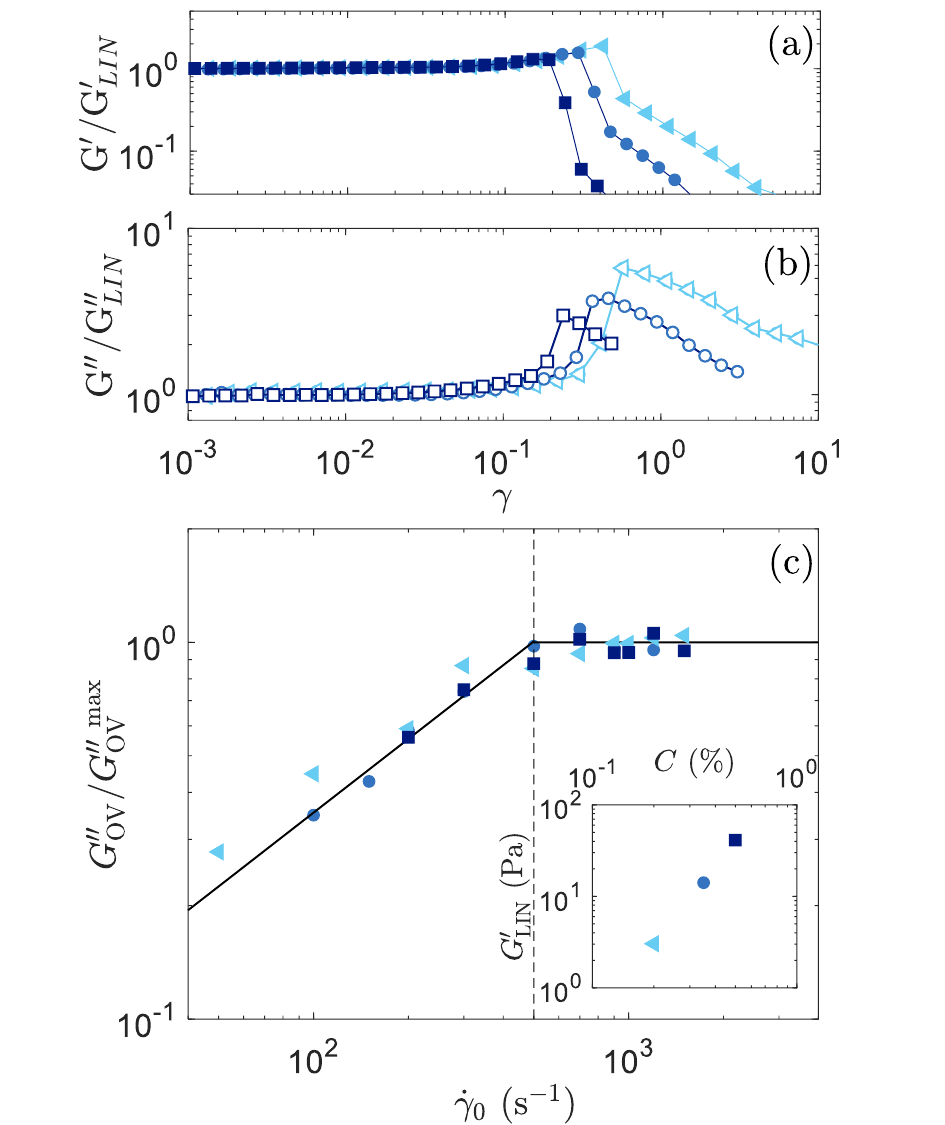}
    \centering
    \caption{Effect of carrageenan concentration ($C_w = 0.5~\%~(w/w)$ -- square, $C_w = 0.35~\%~(w/w)$ -- circle, and $C_w = 0.2~\%~(w/w)$ -- triangle) on the non-linear properties of fluid gels with $I = 40~\rm{mM}$. (a,b) Rescaled storage and loss moduli for the gel obtained with $\dot{\gamma}_0 = 1000~\rm{s}^{-1}$. (c) Rescaled $G^{\prime\prime}$ overshoot as function of the shear rate. Inset displays the evolution of the storage modulus for the fluid gels ($\dot{\gamma}_0 = 1000~\rm{s}^{-1}$) as function of carrageenan concentration.
    }
    \label{figsupp:conc}
\end{figure}

Fig.~\ref{figsupp:conc} shows the non-linear properties of $\kappa$-carrageenan gels at $I = 40~\rm{mM}$ for three carrageenan concentrations. In Fig.~\ref{figsupp:conc}(c), the critical shear rate controlling the evolution of the $G^{\prime\prime}$ overshoot remains constant across concentrations, so the x-axis does not require rescaling, unlike Fig.~3(b) in the main text for different ionic strengths. Only the amplitude of $G^{\prime\prime}_{OV}$ on the y-axis needs rescaling, as it varies with concentration, as shown in Fig.~\ref{figsupp:conc}(b). 
Fig.~\ref{figsupp:conc}(a) shows that the strain-hardening behavior observed across different volume fractions obeys a single master curve when rescaled by the linear elastic modulus $G^{\prime}_{\text{LIN}}$. In particulate colloidal gels, such master curves have been linked to a similar chemical length $d_b$, which dictates the elasticity of the network backbone~\cite{gisler1999strain}. For carrageenan gels, this scaling suggests that the backbone rigidity remains constant across volume fractions, explaining the conservation of the energy density $U/\delta^3$ as indicated by the independence of $\dot{\gamma}^*$ from concentration [Fig.~\ref{figsupp:conc}(c)].

\subsection{Rheology in the Presence of Tracers}
\label{sec:tracers}
As described in the Experimental Section, $1~\%~(w/w)$ skim milk powder is added to a $\kappa$-carrageenan dispersion containing 20~mM KCl to serve as tracers for confocal imaging. Milk at its natural concentration contains approximately 10~\% solids and 100~mM salts; thus, adding 1~\% milk powder is expected to increase the ionic strength by roughly 10~mM.

\begin{figure}[t!]
    \includegraphics[scale=0.53, clip=true, trim=0mm 0mm 0mm 0mm]{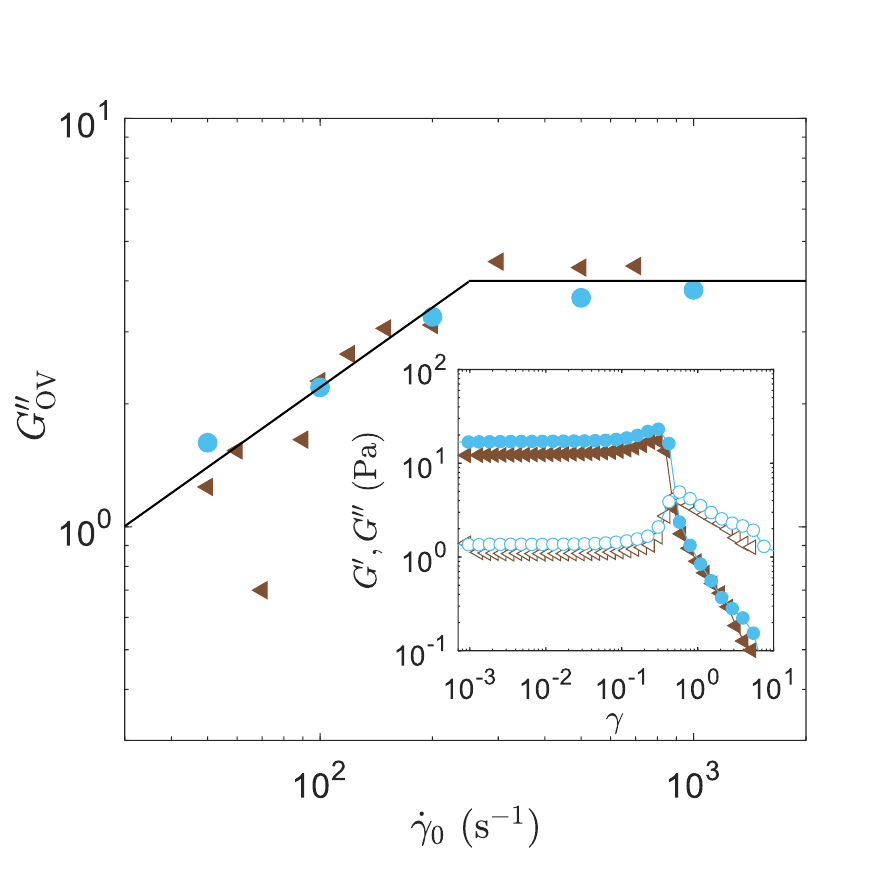}
    \centering
    \caption{Magnitude of $G^{\prime\prime}$ overshoot measured during strain amplitude sweep as function of the preshear rate $\dot{\gamma}_0$ applied during cooling. Triangles show carrageenan dispersion with $30~\rm{mM}$ KCl, while circles indicate $20~\rm{mM}$ KCl + $1~\%$ (w/w) milk powder. Inset displays the strain amplitude sweep for these two conditions for the gels formed at $\dot{\gamma}_0 = 500~\rm{s}^{-1}$. 
    }
    \label{figsupp:microscopy}
\end{figure}

Fig.~\ref{figsupp:microscopy} shows the amplitude of the $G^{\prime\prime}$ overshoot as a function of the cooling shear rate $\dot{\gamma}_0$, as in Fig.~3 of the main text, for the 20~mM dispersion seeded with milk (blue circles) and the 30~mM dispersion (brown triangles). Both dispersions exhibit a critical shear rate $\dot{\gamma}^* \approx 250~\rm{s^{-1}}$, confirming that this rate is governed by ionic strength and that the addition of 1~\% milk increases it by approximately 10~mM. The inset shows that the strain sweeps measured at $\dot{\gamma}_0 = 500~\rm{s^{-1}}$ are superimposed, further demonstrating that the rheological behavior of the two dispersions is identical.

\section{Taylor–Couette Instability}
\label{sec:taylor}
The inset in Fig.~1(a) shows that the shear viscosity of carrageenan dispersions at high temperature remains constant, indicating Newtonian behavior as expected in the dilute regime. At $\dot{\gamma} \approx 700~\mathrm{s^{-1}}$, the viscosity increases, which can be attributed to the onset of the Taylor–Couette instability and the transition toward turbulent flow. This instability arises from inertial effects and the formation of Taylor vortices, typically observed in low-viscosity fluids at high rotational speeds~\cite{Pakdel1996}.

The occurrence of this instability can be anticipated from the Taylor number~\cite{Pakdel1996}:
\begin{equation}
Ta = 2 \left( \frac{d}{r_i} \right) Re^2
\end{equation}
where $r_i$ is the radius of the inner cylinder, $d$ the gap width, and $Re = \Omega \rho r_i^2 / \eta$ the Reynolds number, with $\Omega$, $\rho$, and $\eta$ being the angular velocity, density, and dynamic viscosity of the fluid, respectively.

Assuming the density equals that of water and taking $\eta = 4.5~\mathrm{mPa{\cdot}s}$ [inset in Fig.~1(a)], the deviation observed at $\dot{\gamma} \approx 700~\mathrm{s^{-1}}$ corresponds to $Ta = 9 \times 10^5$, well within the Taylor vortex regime. This confirms the transition from laminar to unstable flow at high shear rates. The critical shear rate governing the microstructural scenario evidenced in the main text lies between $20$ and $875~\mathrm{s^{-1}}$ depending on the ionic strength, indicating that the results reported here are not affected by this instability. Moreover, these instabilities are expected to rapidly diminish near the gelation temperature as viscosity increases with polymer structuring, consistent with the reproducibility of the results and scaling laws reported.

\bibliography{library}